\documentclass[nopreprintline]{elsarticle}

\usepackage{lineno,hyperref}
\usepackage{longtable}
\usepackage{graphicx}
\usepackage{graphicx}

\modulolinenumbers[5]

\journal{}

\usepackage{numcompress}

\begin{document}

\begin{frontmatter}

%%\title{Elsevier \LaTeX\ template\tnoteref{mytitlenote}}
\title{Transfer Learning and SpecAugment applied to SSVEP Based BCI Classification}
%\tnotetext[mytitlenote]{Fully documented templates are available in the elsarticle package on %\href{http://www.ctan.org/tex-archive/macros/latex/contrib/elsarticle}{CTAN}.}

\author{Pedro R. A. S. Bassi*, Willian Rampazzo and Romis Attux}
%%\address{Radarweg 29, Amsterdam}
%%\fntext[myfootnote]{Since 1880.}

%%\author[mymainaddress,mysecondaryaddress]
%%\ead[url]{www.elsevier.com}

%%\author[mysecondaryaddress]{Global Customer Service\corref{mycorrespondingauthor}}
\cortext[mycorrespondingauthor]{Department of Computer Engineering and Industrial Automation, School of Electrical and Computer Engineering, University of Campinas - UNICAMP. 13083-970, Campinas, SP, Brazil. E-Mail: p157007@dac.unicamp.br. Acknowledgements: This work was partially supported by CNPq (process 308811/2019-4) and CAPES.}
%%\ead{support@elsevier.com}

%%\address[mymainaddress]{1600 John F Kennedy Boulevard, Philadelphia}
%%\address[mysecondaryaddress]{360 Park Avenue South, New York}

\begin{abstract}

Objective:
We used deep convolutional neural networks (DCNNs) to classify electroencephalography (EEG) signals in a steady-state visually evoked potentials (SSVEP) based single-channel brain-computer interface (BCI), which does not require calibration on the user.

Methods:
EEG signals were converted to spectrograms and served as input to train DCNNs using the transfer learning technique. We also modified and applied a data augmentation method, SpecAugment, generally employed for speech recognition. Furthermore, for comparison purposes, we classified the SSVEP dataset using Support-vector machines (SVMs) and Filter Bank canonical correlation analysis (FBCCA).

Results:
Excluding the evaluated user's data from the fine-tuning process, we reached 82.2\% mean test accuracy and 0.825 mean F1-Score on 35 subjects from an open dataset, using a small data length (0.5 s), only one electrode (Oz) and the DCNN with transfer learning, window slicing (WS) and SpecAugment's time masks.

Conclusion: 
The DCNN results surpassed SVM and FBCCA performances, using a single electrode and a small data length. Transfer learning provided minimal accuracy change, but made training faster. SpecAugment created a small performance improvement and was successfully combined with WS, yielding higher accuracies.

Significance:
We present a new methodology to solve the problem of SSVEP classification using DCNNs. We also modified a speech recognition data augmentation technique and applied it to the context of BCIs. The presented methodology surpassed performances obtained with FBCCA and SVMs (more traditional SSVEP classification methods) in BCIs with small data lengths and one electrode. This type of BCI can be used to develop small and fast systems.

\end{abstract}


\end{frontmatter}

%%\linenumbers
\section{Introduction}

In this study, we propose the use of deep convolutional neural networks (DCNNs) as classifiers in brain-computer interfaces (BCIs) based on steady-state visually evoked potentials (SSVEP). Deep neural networks (DNNs) perform very well when trained on a large amount of data \cite{imagenet}, but large SSVEP datasets are not commonly available for open use. Our way to overcome this problem was to employ data augmentation and transfer learning techniques to train the DNNs, as both are known to improve the performances of DNNs on smaller datasets \cite{goodfellow2016}.

We started with an open SSVEP dataset \cite{dataset}, which we consider to be large in comparison with other open databases. The electroencephalography (EEG) signals where transformed into images, specifically spectrograms, using the short-time Fourier transform (STFT). By doing so, we take advantage of the ability of convolutional DNN in classifying images, which is well documented \cite{imagenet}.

The neural network used in this study \cite{vgg} is a DCNN based on the original VGG \cite{vggOriginal}. While the original VGG was trained on the ImageNet dataset \cite{imagenet}, the VGGish DCCN we used was trained for classifying spectrograms from AudioSet \cite{AudioSet}, a very large dataset composed of natural and artificial sounds. We decided to investigate this DCNN as our pretrained network for transfer learning, due to the potential task similarity between classifying spectrograms from sounds and classifying spectrograms from SSVEP signals. The structure of this network is found in Table \ref{tab:vggish}. The modified version, fine-tuned for SSVEP classification, is available in Table \ref{tab:ourvgg}.

We trained the network to classify EEG signals from subjects focusing on a 12 Hz or a 15 Hz flickering visual stimulus, using only information from the Oz electrode, thus, simulating a single-channel BCI. We did not include the data from the test subject in the training or validation datasets in order to establish a scenario in which the BCI final user does not influence the classifier adjust. 

During the training period, to avoid overfitting the DNN, we used data augmentation, consisting of window slicing \cite{janelamento} and the technique called SpecAugment \cite{SpecAugment}.  SpecAugment 
very good results in improving the performances of DNNs in the task of speech recognition using spectrograms. Using it, the authors of \cite{SpecAugment} were able to achieve state-of-the art results on the datasets LibriSpeech 960h and Switchboard 300h. Using a LAS network and without the use of a LM, the Switchboard word error rate was reduced by 4\% with the augmentation technique \cite{SpecAugment}. Furthermore, the current (January 2021) LibriSpeech test-other best performing model uses, among other techniques, SpecAugment \cite{switchLeader}.

We compared our results to those of classification using Filter Bank canonical correlation analysis (FBCCA) (the classification technique used by the authors of the dataset that we are using, which outperformed standard CCA in a similar dataset \cite{FBCCA}), DCNN without transfer learning and support vector machines (SVM, a common SSVEP classification method \cite{SVMEx}, \cite{SVMExComparison}).

This paper is divided according to the following way: in section 2, we discuss related studies; in section 3 we present the SSVEP paradigm and the dataset; in section 4, we discuss data augmentation and the techniques we applied (SpecAugment \cite{SpecAugment} and window slicing \cite{janelamento}); in section 5, we introduce the transfer learning and the VGG network \cite{vgg}; in section 6 we describe how we processed the dataset, created our network and trained it; in section 7, we present the results, analyzing the use of different augmentation techniques and comparing the DNN performance to alternative classification methods. Finally, in section 8, we present our conclusions.

\section{Related Work}

Although not the most common SSVEP classification method, convolutional neural networks have been used to analyze SSVEP based BCIs (for example, in \cite{MultiChannelCNN}, \cite{SingleChannelCNN} and \cite{TimeAndFreqCNN}). 

In \cite{MultiChannelCNN} the authors used a SSVEP based BCI to control an exoskeleton in an ambulatory environment. They utilized data from 8 electrodes and converted their signals to the frequency domain, using FFT. For classification the authors employed a convolutional neural network, whose input consisted of a two-dimensional matrix, with 8 rows, each one corresponding to the data generated by one electrode. They achieved accuracies above 94\%, using a data length of 2 s, and surpassed CCA performances.

In \cite{SingleChannelCNN} the authors applied a 1 dimensional DCNN to create a virtual keyboard, using a single-channel SSVEP based BCI. They also created a 3D printed headset, containing only 2 electrodes, which appears to be light and practical, allowing fast self-applicability. They achieved high accuracy, above 97\%, with a 2 s data length, and close to 70\% with a 0.5 s window size. Their CNN results also surpassed CCA.

Finally, the authors of \cite{TimeAndFreqCNN} created a neural network that takes two inputs, a time signal and its FFT (which is analyzed by a CNN). They used a multi-channel approach in the frequency-based input analysis (similar to \cite{MultiChannelCNN}). Their DNN results surpassed CCA when the subjects using the BCI were moving.

Some key differences between this study and the studies mentioned above are: We did not use the test subject data for training, therefore, we simulated a BCI that does not require calibration on the final user. Preliminary testes showed that this makes the classification problem much more difficult for neural networks, as it increases overfitting tendency. We used STFT and not FFT. The authors of \cite{TimeAndFreqCNN} used two separate inputs to feed the DNN time and frequency information, using STFT generated spectrograms, we combined this information into a single input, creating an image. We utilized larger DNNs, in the aforementioned studies, 5 layers or less were used. As deeper architectures are more prone to overfitting on relatively small datasets \cite{goodfellow2016}, we suggested the usage of transfer learning and SpecAugment.

\section{BCI and SSVEP}
\subsection{Fundamentals of SSVEP based BCIs}

A steady-state visually evoked potentials (SSVEP) based brain-computer interface (BCI) utilizes a visual stimulation device, like a computer monitor, showing patterns flickering in certain frequencies. When the user focuses his/her attention on one pattern, he/she sends a command to the system: different frequencies allow different commands \cite{beverina}.

The visual oscillations engender brain activity oscillations (SSVEP) in the same frequency, and their harmonics can be detected with EEG, particularly on the visual cortex. The process is non-invasive, making it a good alternative to create BCIs \cite{galloway1990}. 

In order to identify the oscillations in the EEG signals, we need a classifier. In our work, this role will be played by a deep neural network. Moreover, we want to be able to use the BCI in users that did not take part in its design, hence we do not include the test subject's data in the training set of the DNN. This makes the problem more challenging, increasing the need for regularization and other techniques to avoid overfitting.

\subsection{The dataset}
The authors of the dataset used in this work \cite{dataset} built it to evaluate a virtual keyboard consisting of a computer display showing 40 visual flickers, with different frequencies and phases, corresponding to different letters. The decision to use this dataset was due to its size and availability. It consists of data from 35 subjects, of which 8 subjects are experienced in using BCIs, and 27 subjects do not have any prior experience in using BCIs. The data was recorded with a 64-channel whole head electroencephalography (EEG) in 40 different stimulation frequencies, ranging from 8 to 15.8 Hz, with an interval of 0.2 Hz, while adjacent frequencies had a 0.5 $\pi$ phase difference. Each subject observed the stimuli in six blocks of 40 trials, one for each frequency. Each trial lasted 5 seconds. More details about the dataset, as well as image examples, can be found in \cite{dataset}.

From this dataset, we analyzed the frequencies of 12 and 15 Hz, and only the Oz electrode (based in the international 10-20 system), as it is placed on the region of the visual cortex. In a previous study \cite{bassi_rampazzo_attux_2019}, our group had a better classification performance using this electrode in comparison with other combinations of electrodes.  In that study deep neural networks were trained to classify SSVEP spectrograms, we compared performances when using only the images generated with the Oz electrode and when using images created with different combinations of electrodes. When using many electrodes, the spectrograms generated by their signals were individually produced and processed by the neural networks (in a single-channel approach). Using only the Oz electrode signals improved mean accuracies by about 10\% in SVMs and DNNs (compared to utilizing the electrodes O1, O2, Oz and POz). 

\section{Data augmentation}
\subsection{Fundamentals of data augmentation}

When the neural network inputs become larger, like 224x224 images, the DNN input space becomes high dimensional. This leads to the problem known as ``Curse of dimensionality", and, in order to create a good statistical model of the input and label distributions, our neural network will need more data \cite{Curse}.

Another problem is that, when neural networks become deeper and with more trainable parameters, their ability to model a given data distribution increases. If we do not have enough data, the DNN will learn slight variations and noise in the training dataset, which are exclusive to that database and do not reflect the task the network is trying to solve. This is known as overfitting, and will make the network perform badly in the test dataset \cite{goodfellow2016}.

One way to avoid this problems is to use more training data, giving the network a better statistical representation of the analyzed task. But, sometimes, like in SSVEP classification, large datasets are not available, in which cases we may use data augmentation.

Data augmentation consists in generating more training data from the available training dataset \cite{goodfellow2016}. There are many techniques to accomplish this goal, generally depending on the neural network task. For example, in pattern recognition it is common to rotate or translate the original image, generating new ones.

\subsection{Window slicing}

The first augmentation technique we used in our study was window slicing (WS) \cite{janelamento}, which was also used to define our data length. Small lengths simulate faster BCIs. Our original EEG data consisted of a time series with a duration of 5 seconds, from which we took slices of a fixed size (the data length/window size). The first slice starts at the beginning of the time series, and, for each subsequent slice, we add a fixed displacement considering the start position from the previous slice. The process stops when a slice does not entirely fit in the time series anymore. With this procedure, for each original time series, we create many smaller slices, used as inputs to our neural network, with the same desired output as the original time series taken as the source for the slices. Unlike SpecAugment, window slicing was already used in SSVEP studies (e.g.,  \cite{MultiChannelCNN}).

\subsection{SpecAugment}
One of the areas DCNNs have prominence is image classification. To be able to benefit from this, the EEG signals from the dataset were converted into spectrograms. In addition to that, the process of generating spectrograms creates another opportunity: spectrograms are largely applied to audio classification, as using them we can fully take advantage of two-dimensional convolutional structures in DNNs, similarly as can be done in image classification, obtaining high performances. For example, the model with the best performance in the LibriSpeech test-other speech recognition dataset today (January 2021) utilizes spectrograms and convolutions \cite{switchLeader}. With this in mind, we hypothesized that techniques used for augmenting audio and speech datasets might also be useful for our study. 

To test this new possibility, we used a technique called SpecAugment \cite{SpecAugment}, originally created for data augmentation in speech recognition. It is a modern augmentation method successfully used with DNNs, achieving state-of-art performance in the datasets LibriSpeech 960h and Switchboard 300h \cite{SpecAugment}. The original augmentation strategy consists of three parts \cite{SpecAugment}: time warping, frequency masking, and time masking. Time warping consists in using the TensorFlow sparse\_image\_warp function to warp a random point in the horizontal line passing through the center of the spectrogram to the right or left, by a random distance. Frequency masking consists of masking random frequency channels in the spectrogram (rows). One mask can take just one channel or multiple adjacent ones, in a random manner.  At last, time masking consists of masking random time steps (columns) in the image, and, like in frequency masks, each mask can have a random length (taking one or many adjacent columns). In order to mask rows or columns, we can substitute their values for the spectrogram average value. More information about SpecAugment can be found in \cite{SpecAugment}. In this study we made some changes to the technique, as explained in Section \ref{subsec:signal_processing_data_augmentation}.

\section{Transfer learning}

\subsection{Fundamentals of transfer learning}

The dynamics of knowledge transfer refers to the use of adjusted parameters from a model in a given environment, problem, or database to explore a possible improvement in the generalization of another model in another environment, problem or database \cite{pan2009}.

A common approach is to apply the knowledge transfer technique in situations of data scarcity, since the training of deep models tends to overfit in an insufficiently sized dataset \cite{bengio2013}. Taking advantage of model parameters adjusted on sufficiently large datasets can overcome this problem.

Another feasible approach to knowledge transfer can benefit problems in different domains. Fine-tuning the parameters from a model with good generalization in its domain, using the data from a problem on a different domain, may improve the possibility of solving the problem. The adjusted model can benefit from the original representations of a model to quickly generalize from little data in the new domain \cite{goodfellow2016}.

The two most common uses of knowledge transfer are \cite{bengio2013}:

a) Feature extraction models. The output of a model without its fully connected layers represent the features from the samples; thus, this new model behaves as a feature extractor. Using this feature extractor on an existing dataset generates a new dataset of features. The newly generated dataset serves as input to smaller models when compared to the full model, responsible for the classification or regression, depending on the problem. Training those small models are less time consuming when compared to the training time for a full model.

b) Fine-tuning the model. The strategy is to use a new dataset to fine-tune the parameters of a pretrained complete model, through a few more training iterations. The two options for this approach are whether the model will be fully adjusted or only the layers closer to the end of the model will be adjusted. This decision is related to the similarity between the new dataset and the original dataset, initially used to train the model. The first layers of a deep model tend to extract simple and general features (like locating borders on images), but the last layers tend to learn more complex and task related representations. Thus, the more different the two datasets and tasks, the more layers, starting from the last, will need to be fine-tuned in the new dataset. 

To decide how knowledge transfer best fits into a new dataset, at least two factors must be analyzed: the size of the new dataset and its similarity to the original dataset. There are four possibilities, taking into consideration that the layers at the beginning of the model are responsible for detecting generic features and the layers at the end of the model are responsible for detecting specific features in the original dataset \cite{yosinski2014}:

\begin{enumerate}
\item The new dataset does not have enough samples to adjust the model, and the samples are similar to the original dataset. As the amount of data may not be enough to fully adjust the model, and considering the ability of the model layers to recognize features from the new dataset based on its similarity to the original, adjusting the fully-connected layers may be more reasonable and lead to a better performance.

\item The new dataset has enough samples to adjust the model, and the samples are similar to the original dataset. With enough samples and similar to the original dataset, it is possible to adjust the complete model without worrying about overfitting.

\item The new dataset does not have enough samples to adjust the model and, its samples do not relate to the original dataset. This case turns out to be the most complex as there is not enough data to adjust the fully connected layers of the model, and the data is unrelated to the original dataset. An option is to remove the fully connected layers and some layers from the end of the model, responsible for recognizing specific features from the dataset, connecting a new model, e.g., linear or fully connected layers, at the end in an attempt to improve the possibility to adjust it.

\item The new dataset has enough samples to adjust the model, and the samples do not relate to the original dataset. With enough data, it may be suitable to train the whole model from the beginning \cite{goodfellow2016}. Still, it can be beneficial to initialize the model with pretrained parameters as the amount of data in the new dataset would be enough to adjust the model.

\end{enumerate}

When the choice to address a new problem is knowledge transfer, it is crucial to remember that the solution will be tied to the original architecture, reducing the flexibility of the model. The chosen pretrained model may have input size restrictions, for example. Additionally, an important detail is related to the learning rate. It is common to use lower learning rates when the model parameters are undergoing adjustments since the base to use a pretrained model is its relatively good parameters. Another strategy is to train the DNN layer by layer, that means, start adjusting only the last layers, then unfreezing other layers and training again, until you train the whole model (a frozen layer means that its parameters are not being adjusted in training). This allow us to adjust more the last layers, that are more task specific.

The use of knowledge transfer can lead to benefits such as reduced training time and better accuracy. In any case, its use requires a careful analysis of the datasets and a thorough adjustment to make it work as expected.

\subsection{The VGG network trained in AudioSet}
Searching for a network suitable for transfer learning in the task of SSVEP classification we looked for something deep, trained in a large dataset and in a task somewhat related to SSVEP classification. We could not find any network that met this criteria and was originally trained with SSVEP. But, because we were trying to classify spectrograms, we thought that our task could be related to classifying sounds in a noisy environment.

AudioSet \cite{AudioSet} is a dataset that contains over one million 10-seconds human-labeled sound clips from YouTube, with hundreds of classes, covering human made sounds, musical instruments and environmental sounds. Google released CNNs trained in the task of classifying this dataset, having spectrograms as inputs \cite{vgg}. The network called VGG \cite{vgg} presented a very good result and a variant of it can be downloaded pretrained (also called VGGish). So, that was our choice for transfer learning in this study. 

The pretrained network architecture is very similar to the original VGG configuration A \cite{vggOriginal}, a network trained on the dataset ImageNet \cite{imagenet} for image classification. The downloaded network structure can be seen in Table 1. All layers (convolutional and dense) have a rectified linear unit (ReLU) activation function. All convolutional layers are 2D and have 3x3 sized kernels, with 1x1 stride and padding. All max pooling layers are also 2D, with 2x2 kernel, stride 2 and dilatation 1. This architecture was modified to create the network we used in this study. We will talk about the changes in this paper's section 6.2, but we already present our DNN structure here, in table 2 (to facilitate comparison).

\begin{table}[!ht]
\centering
\begin{tabular}{|l|}
\hline
Convolution, 64 channels \\ \hline
Maxpool \\ \hline
Convolution, 128 channels \\ \hline
Maxpool \\ \hline
Convolution, 256 channels \\ \hline
Convolution, 256 channels \\ \hline
Maxpool \\ \hline
Convolution, 512 channels \\ \hline
Convolution, 512 channels \\ \hline
maxpool \\ \hline
Fully connected, 4096 neurons \\ \hline
Fully connected, 4096 neurons \\ \hline
Fully connected, 128 neurons \\ \hline
\end{tabular}
\caption{VGGish network architecture}
\label{tab:vggish}
\end{table}

\begin{table}[!ht]
\centering
\begin{tabular}{|l|}
\hline
Convolution, 64 channels \\ \hline
Maxpool \\ \hline
Convolution, 128 channels \\ \hline
Maxpool \\ \hline
Convolution, 256 channels \\ \hline
Convolution, 256 channels \\ \hline
Maxpool \\ \hline
Convolution, 512 channels \\ \hline
Convolution, 512 channels \\ \hline
Maxpool \\ \hline
Dropout, 50\% \\ \hline
Fully connected, 512 neurons \\ \hline
Dropout, 50\% \\ \hline
Fully connected, 2 neurons \\ \hline
\end{tabular}
\caption{Our DCNN architecture}
\label{tab:ourvgg}
\end{table}

\section{Methodology}
\subsection{Signal preprocessing and data augmentation} \label{subsec:signal_processing_data_augmentation}
We started with the EEG time series. We had 6 trials of 5 seconds per subject per stimulus frequency. The signal sampling rate was 250 Hz (down sampled from a 1000Hz EEG system). So, each trial provided us with 1250 samples. These data, obtained from \cite{dataset}, were already filtered by a notch filter at 50 Hz in the data recording, to remove common powerline noise. We also applied a common average reference (CAR) filter. With 35 subjects, the frequencies of 12 and 15 Hz and the data from electrode Oz, we had a total of 420 time series.

We started signal preprocessing by applying WS. We opted for a window size of 0.5 s. It simulates a fast BCI, while still providing enough frequency resolution in our spectrograms to distinguish between the 12 Hz and 15 Hz frequencies. Furthermore, preliminary tests showed that, when using larger data lengths, FBCCA \cite{FBCCA} produced better accuracies than our neural networks (e.g., about 4\% higher with 4 s windows). Thus, although using a larger window size would generate higher accuracy, it would have little practical value for this study, as a simpler method (FBCCA) would perform better than the methods we propose here. Other studies (\cite{SingleChannelCNN}) also showed that increasing window size benefits CCA more than it benefits CNNs. 

We used two displacement values for window slicing, 0.5 s and 0.1 s. 0.5 s is the maximum displacement that allows the usage of all data, creating 10 windows, without superposition, from each 5 s original signal. Most experiments in this paper use this displacement value, as it minimizes training time and still allows an effective comparison of the different methods tested in this study (e.g., DCNN with SpecAugment, DCNN with time masks, SVM, FBCCA and transfer learning). 0.1 s displacement generated 46 windows per original signal. This value creates a dataset size similar to the 0.5 s displacement followed by time masking.

After WS, we applied a short-time Fourier transform (STFT) to our new series, generating a spectrogram for each signal. This STFT used a rectangular window with length of 125 (0.5 s), hop length of 62 (0.248 s). The STFT output modulus were converted to decibels and normalized between 0 and 1, creating images. To visualize them, we can multiply their values by 255 and have a gray scale image. We also tried Hann and Blackman windows in the STFT, but their results were worse than the rectangular one, probably due to its better frequency resolution.

The images were filtered, removing columns representing frequencies we had no interest in. In SSVEP classification we are searching for signals in the stimuli frequencies (12 and 15 Hz) ans it's harmonics. Thus, in the final spectrograms, we had only frequencies between 10 and 18 Hz, 22 and 26 Hz and between 28 and 32 Hz. The resulting images had size 8x3 (8 rows/height and 3 columns/width), with a frequency resolution of 2 Hz and a time resolution of 0.167 s. An example can be seen in Figure 1.
\begin{figure}[t]
\caption{Example of a spectrogram (after filtering), corresponding to a 12 Hz stimulus. It has a frequency resolution of 2 Hz and a time resolution of 0.167 s.}
\includegraphics[width=0.3\textwidth]{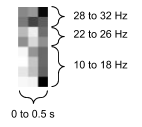}
\centering
\end{figure}

The next step was applying SpecAugment. We made some changes in the original technique. Firstly, we did not use time warp. In the article presenting this method \cite{SpecAugment}, the authors already noted that time warp provided little benefit and was the most computationally expensive operation in SpecAugment. With some early tests we confirmed this, the benefit was marginal and required a significant CPU time. Secondly, our spectrograms were much smaller than the ones used for speech recognition in \cite{SpecAugment}. So, in order to avoid losing too much information, we chose to use only one time mask of 1 column and one frequency mask of 1 line. In this way, in our 8x3 spectrograms, we had only 4 time masking options (one for each row or no mask) and 9 frequency masking options, leaving us with 36 possibilities per image. To avoid repetitions and save training time, we did not use the randomized and online approach suggested in \cite{SpecAugment}, but, instead, we generated and saved all the masking options for each spectrogram before training. 

WS with 0.5 s displacement and SpecAugment yielded us a dataset of 151,200 images. In order to test SpecAugment's benefits we also trained networks without it and networks with time masks only. Time masking provided us a dataset of 16,800 images, also using WS with 0.5 s displacement. Without SpecAugment and utilizing 0.5 s WS displacement, our dataset size was 4,200. Without SpecAugment and using 0.1 s WS displacement, 19,320. With 0.1 s displacement and just time masks our dataset had 77,280 images. 

In this study, we wanted to simulate the use of a BCI in a person who did not participate in the classifier training. Thus, when we choose a test subject, its data is not used in the training or validating datasets. To create these datasets, we removed the test subject data and randomly selected 66.6\% of the remaining spectrograms as the training dataset and 33.3\% as the validating one (but maintaining the balance between classes). The test dataset is created with all spectrograms from the test subject. We note here that SpecAugment was only applied to the training dataset and never to testing or validating. WS was applied to all the datasets (with the same parameters), because we want to use the same data length (0.5 s) for training and testing, more effectively simulating a fast BCI. Therefore, we perform training, testing and validation utilizing only the data generated by window slicing.

To feed the spectrograms to the DNNs, they were resized (using nearest neighbor interpolation) to 96x64 (96 rows/height and 64 columns/width), the pretrained VGG input size. Our SVMs were trained with the same dataset of spectrograms used to train the deep networks. The 8x3 images were reshaped into vectors with length 24 (using PyTorch .view) before being sent to the SVMs, which mapped them to a single output.

\subsection{Training, transfer learning and alternative classification methods}

To create our network, we started from the pretrained VGG, described in section 5.2. But, after some preliminary testing, we decided to maintain the convolutional layers, but to replace the fully connected ones (the last 3 layers) by two new fully connected layers, with random weights, and add dropouts, as shown in Table 2. The first fully-connected layer has ReLU activation functions and the last layer is linear (the model will be trained in PyTorch with cross-entropy loss, which adds a softmax activation to the output). We also made tests without removing layers and just adding new ones to the end of the VGG, or changing the last VGG layer, but their results were worse.

Preliminary tests also shown that the best results were achieved without freezing any layers weight when training. We also tried first freezing the convolutional block, training only the fully connected one, and then training the rest of the network (at once or unfreezing and training layer by layer).

Because we are not using the test subject's data in training or validating, much regularization was needed to avoid overfitting, in form of max pooling, dropout, weight decay and early stopping. We trained one network for each of the 35 subjects as the test one.
 
All the methods in this study were implemented in Python, using a NVidia RTX 3080 graphics processing unit and the library PyTorch. We note that the neural networks did not achieve the maximum number of training epochs, which we will state bellow, because we used early stop.

 Using WS with 0.5 s window displacement, we tested our DCNN with SpecAugment's time and frequency masks, with time masks only, and without this augmentation. The networks were trained with stochastic gradient descent (SGD), momentum of 0.9, learning rate of 0.001, weight decay of 0.01, cross-entropy loss and mini-batches of 128. Training with SpecAugment we utilized early stop with patience of 50 and maximum of 500 epochs, with only time masks or without SpecAugment we used patience of 500 and maximum of 5000 epochs. Training times are dependent on the SpecAugment utilization. Using time and frequency masks (generating our biggest dataset), each epoch lasted about 30 seconds, with only time masks, 3.6 seconds, and without SpecAugment, 1.3 seconds. The number of epochs in training were bigger when using less augmentation. Using SpecAugment, an optimal neural network (according to validation loss) is achieved at about 1,100 s, using time masks, about 730 s, and, without SpecAugment, about 270 s.
 
 Using WS with 0.1 s window displacement, we trained a DNN without SpecAugment and one with time masking. They were also trained with stochastic gradient descent (SGD), momentum of 0.9, learning rate of 0.001, weight decay of 0.01, cross-entropy loss and mini-batches of 128. With time masks, we used early stop with patience of 250 and maximum of 2000 epochs. Without SpecAugment, patience of 200 and maximum of 5000 epochs. Each epoch utilizing time masks took about 15 s and, without SpecAugment, about 6 s. Without SpecAugment, we needed about 2,700 s to obtain an optimal DNN, with time masks, 5,600 s.

 We also used three alternative SSVEP classification methods to provide a good comparison to our results: the DNN shown in table 2 without transfer learning, SVMs and FBCCA. 
 
 We trained the DNN without transfer learning with the same parameters described above (i.e., same loss function, optimizer parameters and batch size), without SpecAugment, maximum of 5000 epochs, early stop with patience of 2000 and WS using 0.5 s displacement. We used a higher patience because, unlike the DNNs that used transfer learning, these networks had a slow start in training, as losses could stay in high a plateau for hundreds of epochs before starting to decrease. Epochs took about 1.3 s, and mean time to obtain an optimal neural network was about 310 s.

SVMs are a commonly used SSVEP classification method. Training them also allowed us to compare the benefits of SpecAugment both in a shallow and a deep model. We decided to use linear kernels because they are simple and commonly used in SSVEP classification (e.g., in \cite{SVMEx} and \cite{SVMExComparison}). Furthermore, as far as we could check, there are no substantial performance increase, when using other SVM kernels. For example, in \cite{SVMExComparison} the authors used SVMs as classifiers in a SSVEP based BCI, which controlled a RF car. They tested four types of kernel, linear, RBF, polynomial and sigmoid, and the mean command recognition rates achieved with them were 90.16\%, 90.29\%, 84.25\% and 90.19\%, respectively. We trained linear SVMs with SGD, momentum of 0.9, learning rate of 0.01, hinge loss with a regularization parameter (c) of 0.01, early stop of 200, maximum number of epochs of 5000, mini-batches of 128 and 0.5 s WS window displacement. Without SpecAugment, epochs took about 0.5 s and the mean time to obtain an optimal SVM was about 100 s. With SpecAugment epochs had a 4.3 s duration and about 660 s were needed to obtain an optimal network. 

Finally, we performed FBCCA SSVEP classification. This is the technique used by the authors of the dataset that we are using \cite{dataset}, and it could surpass CCA performances in a similar database \cite{FBCCA}. We used the same parameters as they \cite{dataset} (M3 filtering method, Nh=5, a=1.25, b=0.25, N=7), which are better explained in \cite{FBCCA}. We classified the signals that we had created by window slicing (without the STFT), therefore, our data length was 0.5 s. We performed FBCCA using only the Oz electrode data (as we did with our neural networks), and also using 9 electrodes (Pz, PO5, PO3, POz, PO4, PO6, O1, Oz, and O2), as the authors in \cite{FBCCA} and \cite{dataset} did. We based our FBCCA implementation on the code available at \cite{FBCCADrone}, which implements the method described in \cite{FBCCA}, using Python.

\section{Results and discussion}

 In table 3 we present our DCNNs trained with transfer learning. The different networks are indicated by the table columns. The table shows the test accuracies (in percentage) and each row represents a different subject taken as the test subject (whose data is not used for training or validation). The last row shows the mean accuracies in the 35 test subjects, for each DCNN. All DNNs in the table were trained with WS using 0.5 s window displacement, except for the networks indicated by the two columns at the right side, which were trained utilizing 0.1 s displacement.

\begin{table}[]
\centering
\caption{Test accuracies for DCNNs with transfer learning and different data augmentation methods. }
\label{tab:my-table}
\begin{tabular}{|r|r|r|r|r|r|}
\hline
\multicolumn{1}{|l|}{Test}     & \multicolumn{1}{l|}{DCNN with}   & \multicolumn{1}{l|}{DCNN without} & \multicolumn{1}{l|}{DCNN with}  & \multicolumn{1}{l|}{DCNN with}    & \multicolumn{1}{l|}{DCNN with}      \\
\multicolumn{1}{|l|}{subject}  & \multicolumn{1}{l|}{SpecAugment} & \multicolumn{1}{l|}{SpecAugment}                       & \multicolumn{1}{l|}{time masks} & \multicolumn{1}{l|}{0.1 s displacement} & \multicolumn{1}{l|}{0.1 s displacement}   \\
\multicolumn{1}{|l|}{}         & \multicolumn{1}{l|}{}            & \multicolumn{1}{l|}{}                                  & \multicolumn{1}{l|}{}           & \multicolumn{1}{l|}{(no SpecAugment)}             & \multicolumn{1}{l|}{and time masks} \\ \hline
1                              & 75.8                             & 72.5                                                   & 69.2                            & 74.5                              & 74.6                                \\ \hline
2                              & 78.3                             & 76.7                                                   & 79.2                            & 80.4                              & 81.7                                \\ \hline
3                              & 90.8                             & 90.8                                                   & 90.8                            & 93.5                              & 93.7                                \\ \hline
4                              & 71.7                             & 70.8                                                   & 75                              & 72.6                              & 72.6                                \\ \hline
5                              & 91.7                             & 91.7                                                   & 91.7                            & 91.7                              & 91.7                                \\ \hline
6                              & 89.2                             & 90                                                     & 89.2                            & 90.9                              & 90.9                                \\ \hline
7                              & 79.2                             & 80                                                     & 79.2                            & 83.2                              & 84.4                                \\ \hline
8                              & 93.3                             & 92.5                                                   & 91.7                            & 92                                & 92.2                                \\ \hline
9                              & 74.2                             & 73.3                                                   & 75                              & 78.3                              & 78.4                                \\ \hline
10                             & 87.5                             & 85                                                     & 85.8                            & 88.8                              & 88.8                                \\ \hline
11                             & 66.7                             & 64.2                                                   & 64.2                            & 63.9                              & 63.2                                \\ \hline
12                             & 65.8                             & 64.2                                                   & 66.7                            & 69.7                              & 70.8                                \\ \hline
13                             & 55.8                             & 56.7                                                   & 57.5                            & 60                                & 60.9                                \\ \hline
14                             & 70                               & 68.3                                                   & 70.8                            & 68.8                              & 69.6                                \\ \hline
15                             & 87.5                             & 88.3                                                   & 91.7                            & 88.4                              & 88                                  \\ \hline
16                             & 70                               & 70.8                                                   & 69.2                            & 75.5                              & 76.3                                \\ \hline
17                             & 75.8                             & 76.7                                                   & 75.8                            & 78.4                              & 79.2                                \\ \hline
18                             & 75.8                             & 77.5                                                   & 78.3                            & 76.3                              & 76.8                                \\ \hline
19                             & 75                               & 72.5                                                   & 74.2                            & 71.7                              & 72.5                                \\ \hline
20                             & 90.8                             & 90                                                     & 90.8                            & 89.9                              & 91.3                                \\ \hline
21                             & 72.5                             & 72.5                                                   & 69.2                            & 74.8                              & 74.3                                \\ \hline
22                             & 98.3                             & 96.7                                                   & 96.7                            & 96.6                              & 96.4                                \\ \hline
23                             & 74.2                             & 66.7                                                   & 70                              & 67.4                              & 67.8                                \\ \hline
24                             & 92.5                             & 92.5                                                   & 93.3                            & 91.8                              & 91.7                                \\ \hline
25                             & 86.7                             & 84.2                                                   & 85.8                            & 84.8                              & 85.3                                \\ \hline
26                             & 91.7                             & 90.8                                                   & 90                              & 92.2                              & 92.4                                \\ \hline
27                             & 89.2                             & 92.5                                                   & 93.3                            & 92                                & 92.6                                \\ \hline
28                             & 86.7                             & 87.5                                                   & 88.3                            & 93.1                              & 93.3                                \\ \hline
29                             & 60                               & 54.2                                                   & 57.5                            & 57.6                              & 58.7                                \\ \hline
30                             & 93.3                             & 94.2                                                   & 94.2                            & 94.7                              & 93.7                                \\ \hline
31                             & 89.2                             & 91.7                                                   & 90.8                            & 91.7                              & 92                                  \\ \hline
32                             & 95                               & 95.8                                                   & 95                              & 97.6                              & 97.8                                \\ \hline
33                             & 70                               & 69.2                                                   & 71.7                            & 72.6                              & 73.4                                \\ \hline
34                             & 78.3                             & 80.8                                                   & 80                              & 76.4                              & 77.9                                \\ \hline
35                             & 86.7                             & 86.7                                                   & 88.3                            & 89.7                              & 90.8                                \\ \hline
\multicolumn{1}{|l|}{Mean}     & 80.8                             & 80.2                                                   & 80.8                            & 81.8                              & 82.2                                \\
\multicolumn{1}{|l|}{accuracy} & \multicolumn{1}{l|}{}            & \multicolumn{1}{l|}{}                                  & \multicolumn{1}{l|}{}           & \multicolumn{1}{l|}{}             & \multicolumn{1}{l|}{}               \\ \hline
\end{tabular}
\end{table}

 In table 4 we show the test accuracies generated by the alternative classification methods, i.e., the DCNN without transfer learning, SVM and FBCCA. All methods utilized WS with 0.5 s window displacement. We note that, in this table, SpecAugment was only used with one of the SVMs. Also, except for the last column (FBCCA with 9 electrodes), every method used only the Oz electrode data.

\begin{table}[]
\centering
\caption{Test accuracies for alternative SSVEP classification methods.}
\label{tab:my-table}
\begin{tabular}{|r|r|r|r|l|l|}
\hline
\multicolumn{1}{|l|}{Test}                              & \multicolumn{1}{l|}{SVM}  & \multicolumn{1}{l|}{SVM with} & \multicolumn{1}{l|}{DCNN without}      & FBCCA & FBCCA             \\
\multicolumn{1}{|l|}{Subject}                           & \multicolumn{1}{l|}{}     & \multicolumn{1}{l|}{SpecAugment}                   & \multicolumn{1}{l|}{transfer learning} &       & with 9 electrodes \\ \hline
1                                                       & 72.5                      & 69.2                                               & 74.2
                     & 75   & 93.3              \\ \hline
2                                                       & 80                        & 79.2                                               & 78.3                                   & 87.5  & 96.7              \\ \hline
3                                                       & 89.2                      & 90                                                 & 90.8                                   & 93.3  & 95.8              \\ \hline
4                                                       & 65.8                      & 67.5                                               & 72.5                                   & 68.3  & 95.8              \\ \hline
5                                                       & 90                        & 90.8                                               & 90                                     & 90.8  & 95.8              \\ \hline
6                                                       & 87.5                      & 85                                                 & 88.3                                   & 86.7  & 94.2              \\ \hline
7                                                       & 79.2                      & 78.3                                               & 80.8                                   & 77.5  & 92.5              \\ \hline
8                                                       & 90.8                      & 90                                                 & 89.2                                   & 85.8  & 90.8              \\ \hline
9                                                       & 76.7                      & 76.7                                               & 70                                     & 70.8  & 83.3              \\ \hline
10                                                      & 84.2                      & 83.3                                               & 85.8                                   & 91.7  & 94.2              \\ \hline
11                                                      & 62.5                      & 61.7                                               & 60.8                                   & 60    & 85                \\ \hline
12                                                      & 63.3                      & 64.2                                               & 62.5                                   & 65    & 95                \\ \hline
13                                                      & 53.3                      & 55                                                 & 57.5                                   & 55    & 95                \\ \hline
14                                                      & 61.7                      & 64.2                                               & 72.5                                   & 68.3  & 95                \\ \hline
15                                                      & 85                        & 85                                                 & 89.2                                   & 80    & 87.5              \\ \hline
16                                                      & 67.5                      & 66.7                                               & 70.8                                   & 71.7  & 80.8              \\ \hline
17                                                      & 77.5                      & 75.8                                               & 75.8                                   & 71.7  & 82.5              \\ \hline
18                                                      & 75.8                      & 76.7                                               & 76.7                                   & 70.8  & 90.8              \\ \hline
19                                                      & 69.2                      & 69.2                                               & 71.7                                   & 63.3  & 69.2              \\ \hline
20                                                      & 90                        & 88.3                                               & 89.2                                   & 87.5  & 84.2              \\ \hline
21                                                      & 72.5                      & 71.7                                               & 75                                     & 70.8  & 85.8              \\ \hline
22                                                      & 96.7                      & 95.8                                               & 97.5                                   & 88.3  & 95                \\ \hline
23                                                      & 70                        & 70                                                 & 70                                     & 67.5  & 91.7              \\ \hline
24                                                      & 83.3                      & 82.5                                               & 94.2                                   & 90.8  & 96.7              \\ \hline
25                                                      & 83.3                      & 81.7                                               & 82.5                                   & 71.7  & 99.2              \\ \hline
26                                                      & 89.2                      & 86.7                                               & 91.7                                   & 71.7  & 95.8              \\ \hline
27                                                      & 90.8                      & 90.8                                               & 90                                     & 92.5  & 93.3              \\ \hline
28                                                      & 89.2                      & 88.3                                               & 87.5                                   & 88.3  & 95.8              \\ \hline
29                                                      & 59.2                      & 60                                                 & 59.2                                   & 57.5  & 81.7              \\ \hline
30                                                      & 91.7                      & 92.5                                               & 94.2                                   & 89.2  & 94.2              \\ \hline
31                                                      & 88.3                      & 87.5                                               & 91.7                                   & 82.5  & 98.3              \\ \hline
32                                                      & 95.8                      & 95                                                 & 95                                     & 95.8  & 93.3              \\ \hline
33                                                      & 66.7                      & 68.3                                               & 70.8                                   & 68.3  & 81.7              \\ \hline
34                                                      & 80.8                      & 82.5                                               & 80                                     & 65.8  & 96.7              \\ \hline
35                                                      & 86.7                      & 89.2                                               & 85.8                                   & 76.7  & 91.7              \\ \hline
Mean accuracy        & 79                        & 78.8                                               & 80.3                                   & 77.1  & 91.1              \\ \hline
\end{tabular}
\end{table}

Analyzing our DCNNs' mean F1-Scores, we observe that the highest value, 0.825, was achieved by our DCNN with time masks and 0.1 s window displacement in WS. The second best, 0.819, was achieved by the DCNN with 0.1 s window displacement and without SpecAugment. Analyzing the networks that we trained with the 0.5 s window displacement, we observe that the DCNN with SpecAugment achieved 0.814 mean F1-Score; the DCNN with time masks, 0.811; the DCNN without SpecAugment achieved 0.806 and the DCNN without transfer learning, 0.805. The SVMs achieved mean F1-Scores of 0.794 with SpecAugment and 0.793 without it. FBCCA with the Oz electrode provided 0.754 mean F1-Score, and with 9 electrodes, 0.908. It is interesting to notice that SpecAugment (and just time masks) created a slightly higher improvement to mean F1-Score than to mean accuracy.

Comparing the augmentation methods of SpecAugment and WS, we observe that the DCNN using SpecAugment and WS with 0.5 s displacement could not surpass the mean accuracy of the network without SpecAugment and using WS with 0.1 s window displacement (even though the dataset sizes are similar in both cases). But we also see that the two augmentation methods can be successfully combined, as the DCNN with time masks and WS using 0.1 s window displacement achieved the best mean accuracy and F1-Score in this study (using one electrode).

In table 3, analyzing the networks that used 0.5 s window displacement, we observe that our DCNN with SpecAugment's time and frequency masking, and the one with just time masks, had the same mean accuracy, which was higher than the DCNN without SpecAugment. In table 4 we can also observe that this augmentation has a much smaller effect on the shallow and linear SVM (actually, its mean accuracy is reduced by a very small value, 0.2\%). 

We observe that all DNNs' mean accuracies were superior than the SVMs'. We also note that the mean performance difference between the DCNN with transfer learning and without it (both without SpecAugment) is almost null. But transfer learning presents two clear advantages: it made training faster, with about 15\% less time to obtain an optimal DNN; it also made the training and validation losses start to decrease in the first couple epochs, whereas, without it, the losses could stay stagnated, in an elevated value, for hundreds of epochs, before finally starting to drop (creating a more unpredictable training process, which also required more patience in early stopping).

Table 4 shows that FBCCA's mean accuracy is worse than the neural networks' when it also only analyzes the Oz electrode data. It has much better results with a multi-channel approach, using 9 electrodes (Pz, PO5, PO3, POz, PO4, PO6, O1, Oz, and O2). This is expected, as FBCCA is a multi-channel method, which highly improves accuracy when using data from many electrodes. The DCNNs we proposed here are a single-channel approach, as they take a single spectrogram as input. In a previous study \cite{bassi_rampazzo_attux_2019}, we could not obtain superior results training neural networks with data from many electrodes (as explained in section 3.2). But, in that study, although we created spectrograms using signals from many electrodes, these images were created independently (each STFT was performed taking a signal from a single electrode, and the generated spectrogram was added to the dataset). The DCNNs also analyzed the spectrograms independently (one at a time), thus, they also used a single-channel approach, not combining the information provided by multiple electrodes, like FBCCA does. 

Other result worth noting is that training our network with frozen weights, and then unfreezing layers and retraining, did not improve our accuracies. 

Finally, we should state that, when comparing our results to other studies, it is important to pay attention if they also do not use the test subject's data for training. This approach decreases DNN accuracy, in exchange of simulating a BCI that does not require calibration on the final user.

\section{Conclusion}

With this study we could observe that deep convolutional neural networks can improve the accuracy of a single-channel SSVEP based BCI, even without training in the test subject, when using small data lengths, like 0.5 s (in comparison to SVMs and FBCCA). Our DCNN with time masks, WS using 0.1 s window displacement, and transfer learning had a mean accuracy that was 3.2\% higher than the SVM's and 5.1\% higher than the FBCCA's. Improving performances on small data lengths is an important factor to create faster BCIs, and using only one electrode can allow the creation of smaller, more portable and potentially less expensive devices, like the 3D printed headset shown in \cite{SingleChannelCNN}, which allows a simple self-application.

The SSVEP classification with DCNNs that we proposed here is a single-channel approach. As shown in the last section, a multi-channel approach, using FBCCA, can effectively utilize data from multiple electrodes and produce better accuracy. In a future study, we intend to modify the methods proposed in this paper to also analyze their effectiveness in a multi-channel approach.

Transfer learning almost did not change performances in this study (mean accuracy decreased by 0.1\% and mean F1-Score increased by 0.001). However, it allowed for a faster (about 15\%) and more predictable training procedure, as without it the training and validation losses could take hundreds of epochs to start improving. The absence of accuracy improvement with transfer learning and the ineffectiveness of freezing layers during training might indicate that audio classification is not a task similar enough to EEG classification to fully take advantage of the transfer learning technique. If this is the case, pre-training on very large open SSVEP datasets could lead to better performances.

SpecAugment improved mean accuracies by 0.6\% and mean F1-Score by 0.008 (in comparison to the DCNN without SpecAugment, both using 0.5 s WS displacement). By using a smaller (0.1 s) window displacement in WS we surpassed SpecAugment's accuracies (by 1\%, considering that the dataset size with the smaller displacement is similar to the one using time masks and 0.5 s displacement). However, time masks combined with the smaller displacement provided the best mean accuracies and F1-Scores with one electrode in this study (82.2\% and 0.825, respectively). This shows that SpecAugment can be used with other types of augmentation and improve their performances. As expected, SpecAugment had a much smaller effect on the shallow SVM, actually reducing its accuracy by 0.2\%. 

 Using SpecAugment's time and frequency masks or just time masks provided almost the same performance. We think that the main reason for this is the fact that our spectrograms are very small and frequency information is very important for our classification task. Therefore, although frequency masking may reduce overfitting, it may also remove substantial information from our images. 

This study shows that deep neural networks can improve the performance of SSVEP based BCIs that use small data lengths and a single electrode, and we intend to expand this favorable results to the multiple electrodes case in future works. Also, we think that efforts in creating larger and open SSVEP classification datasets will improve DCNNs' results even more, allowing better generalization and more effective transfer learning. 

%\section*{References}

\bibliography{mybibfile}

\begin{thebibliography}{23}
\providecommand{\natexlab}[1]{#1}
\providecommand{\url}[1]{\texttt{#1}}
\providecommand{\href}[2]{#2}
\providecommand{\path}[1]{#1}
\providecommand{\eprint}[1]{\href{http://arxiv.org/abs/#1}{\path{#1}}}
\providecommand{\DOIprefix}{doi:}
\providecommand{\ArXivprefix}{arXiv:}
\providecommand{\URLprefix}{URL: }
\providecommand{\Pubmedprefix}{pmid:}
\providecommand{\doi}[1]{\href{http://dx.doi.org/#1}{\path{#1}}}
\providecommand{\Pubmed}[1]{\href{pmid:#1}{\path{#1}}}
\providecommand{\BIBand}{and}
\providecommand{\bibinfo}[2]{#2}
\ifx\xfnm\undefined \def\xfnm[#1]{\unskip,\space#1}\fi
%Type = Inproceedings
\bibitem[{Deng et~al.(2009)Deng, Dong, Socher, Li, Li and Fei-Fei}]{imagenet}
\bibinfo{author}{Deng\xfnm[ J.]}, \bibinfo{author}{Dong\xfnm[ W.]},
  \bibinfo{author}{Socher\xfnm[ R.]}, \bibinfo{author}{Li\xfnm[ L.J.]},
  \bibinfo{author}{Li\xfnm[ K.]}, \bibinfo{author}{Fei-Fei\xfnm[ L.]}.
\newblock \bibinfo{title}{{ImageNet: A Large-Scale Hierarchical Image
  Database}}.
\newblock In: \bibinfo{booktitle}{CVPR09}. \bibinfo{year}{2009},.
%Type = Book
\bibitem[{Goodfellow et~al.(2016)Goodfellow, Bengio, Courville and
  Bengio}]{goodfellow2016}
\bibinfo{author}{Goodfellow\xfnm[ I.]}, \bibinfo{author}{Bengio\xfnm[ Y.]},
  \bibinfo{author}{Courville\xfnm[ A.]}, \bibinfo{author}{Bengio\xfnm[ Y.]}.
\newblock \bibinfo{title}{Deep learning}; vol.~\bibinfo{volume}{1}.
\newblock \bibinfo{publisher}{MIT press Cambridge}; \bibinfo{year}{2016}.
%Type = Article
\bibitem[{{Wang} et~al.(2017){Wang}, {Chen}, {Gao} and {Gao}}]{dataset}
\bibinfo{author}{{Wang}\xfnm[ Y.]}, \bibinfo{author}{{Chen}\xfnm[ X.]},
  \bibinfo{author}{{Gao}\xfnm[ X.]}, \bibinfo{author}{{Gao}\xfnm[ S.]}.
\newblock \bibinfo{title}{A benchmark dataset for ssvep-based brain–computer
  interfaces}.
\newblock \bibinfo{journal}{IEEE Transactions on Neural Systems and
  Rehabilitation Engineering}
  \bibinfo{year}{2017};\bibinfo{volume}{25}(\bibinfo{number}{10}):\bibinfo{pages}{1746--1752}.
%Type = Incollection
\bibitem[{Hershey et~al.(2017)Hershey, Chaudhuri, Ellis, Gemmeke, Jansen, Moore
  et~al.}]{vgg}
\bibinfo{author}{Hershey\xfnm[ S.]}, \bibinfo{author}{Chaudhuri\xfnm[ S.]},
  \bibinfo{author}{Ellis\xfnm[ D.P.W.]}, \bibinfo{author}{Gemmeke\xfnm[ J.F.]},
  \bibinfo{author}{Jansen\xfnm[ A.]}, \bibinfo{author}{Moore\xfnm[ C.]}, et~al.
\newblock \bibinfo{title}{Cnn architectures for large-scale audio
  classification}.
\newblock In: \bibinfo{booktitle}{International Conference on Acoustics, Speech
  and Signal Processing (ICASSP)}. \bibinfo{year}{2017},\URLprefix
  \url{https://arxiv.org/abs/1609.09430}.
%Type = Misc
\bibitem[{Simonyan and Zisserman(2014)}]{vggOriginal}
\bibinfo{author}{Simonyan\xfnm[ K.]}, \bibinfo{author}{Zisserman\xfnm[ A.]}.
\newblock \bibinfo{title}{Very deep convolutional networks for large-scale
  image recognition}.
\newblock \bibinfo{year}{2014}.
\newblock \href{http://arxiv.org/abs/1409.1556}{\tt arXiv:1409.1556}.
%Type = Inproceedings
\bibitem[{Gemmeke et~al.(2017)Gemmeke, Ellis, Freedman, Jansen, Lawrence, Moore
  et~al.}]{AudioSet}
\bibinfo{author}{Gemmeke\xfnm[ J.F.]}, \bibinfo{author}{Ellis\xfnm[ D.P.W.]},
  \bibinfo{author}{Freedman\xfnm[ D.]}, \bibinfo{author}{Jansen\xfnm[ A.]},
  \bibinfo{author}{Lawrence\xfnm[ W.]}, \bibinfo{author}{Moore\xfnm[ R.C.]},
  et~al.
\newblock \bibinfo{title}{Audio set: An ontology and human-labeled dataset for
  audio events}.
\newblock In: \bibinfo{booktitle}{Proc. IEEE ICASSP 2017}.
  \bibinfo{address}{New Orleans, LA}; \bibinfo{year}{2017},.
%Type = Misc
\bibitem[{Cui et~al.(2016)Cui, Chen and Chen}]{janelamento}
\bibinfo{author}{Cui\xfnm[ Z.]}, \bibinfo{author}{Chen\xfnm[ W.]},
  \bibinfo{author}{Chen\xfnm[ Y.]}.
\newblock \bibinfo{title}{Multi-scale convolutional neural networks for time
  series classification}.
\newblock \bibinfo{year}{2016}.
\newblock \href{http://arxiv.org/abs/1603.06995}{\tt arXiv:1603.06995}.
%Type = Article
\bibitem[{Park et~al.(2019)Park, Chan, Zhang, Chiu, Zoph, Cubuk
  et~al.}]{SpecAugment}
\bibinfo{author}{Park\xfnm[ D.S.]}, \bibinfo{author}{Chan\xfnm[ W.]},
  \bibinfo{author}{Zhang\xfnm[ Y.]}, \bibinfo{author}{Chiu\xfnm[ C.C.]},
  \bibinfo{author}{Zoph\xfnm[ B.]}, \bibinfo{author}{Cubuk\xfnm[ E.D.]}, et~al.
\newblock \bibinfo{title}{Specaugment: A simple data augmentation method for
  automatic speech recognition}.
\newblock \bibinfo{journal}{Interspeech 2019} \bibinfo{year}{2019};\URLprefix
  \url{http://dx.doi.org/10.21437/Interspeech.2019-2680}.
  \DOIprefix\doi{10.21437/interspeech.2019-2680}.
%Type = Misc
\bibitem[{Zhang et~al.(2020)Zhang, Qin, Park, Han, Chiu, Pang
  et~al.}]{switchLeader}
\bibinfo{author}{Zhang\xfnm[ Y.]}, \bibinfo{author}{Qin\xfnm[ J.]},
  \bibinfo{author}{Park\xfnm[ D.S.]}, \bibinfo{author}{Han\xfnm[ W.]},
  \bibinfo{author}{Chiu\xfnm[ C.C.]}, \bibinfo{author}{Pang\xfnm[ R.]}, et~al.
\newblock \bibinfo{title}{Pushing the limits of semi-supervised learning for
  automatic speech recognition}.
\newblock \bibinfo{year}{2020}.
\newblock \href{http://arxiv.org/abs/2010.10504}{\tt arXiv:2010.10504}.
%Type = Article
\bibitem[{Chen et~al.(2015)Chen, Wang, Gao, Jung and Gao}]{FBCCA}
\bibinfo{author}{Chen\xfnm[ X.]}, \bibinfo{author}{Wang\xfnm[ Y.]},
  \bibinfo{author}{Gao\xfnm[ S.]}, \bibinfo{author}{Jung\xfnm[ T.P.]},
  \bibinfo{author}{Gao\xfnm[ X.]}.
\newblock \bibinfo{title}{Filter bank canonical correlation analysis for
  implementing a high-speed ssvep-based brain–computer interface}.
\newblock \bibinfo{journal}{Journal of neural engineering}
  \bibinfo{year}{2015};\bibinfo{volume}{12}:\bibinfo{pages}{046008}.
\newblock \DOIprefix\doi{10.1088/1741-2560/12/4/046008}.
%Type = Article
\bibitem[{Sutjiadi et~al.(2018)Sutjiadi, Pattiasina and Lim}]{SVMEx}
\bibinfo{author}{Sutjiadi\xfnm[ R.]}, \bibinfo{author}{Pattiasina\xfnm[ T.]},
  \bibinfo{author}{Lim\xfnm[ R.]}.
\newblock \bibinfo{title}{Ssvep-based brain-computer interface for computer
  control application using svm classifier}.
\newblock \bibinfo{journal}{International Journal of Engineering \& Technology}
  \bibinfo{year}{2018};\bibinfo{volume}{7}:\bibinfo{pages}{2722--2728}.
\newblock \DOIprefix\doi{10.14419/ijet.v7i4.16139}.
%Type = Article
\bibitem[{Setiono et~al.(2018)Setiono, Handojo, Intan, Sutjiadi and
  Lim}]{SVMExComparison}
\bibinfo{author}{Setiono\xfnm[ T.]}, \bibinfo{author}{Handojo\xfnm[ A.]},
  \bibinfo{author}{Intan\xfnm[ R.]}, \bibinfo{author}{Sutjiadi\xfnm[ R.]},
  \bibinfo{author}{Lim\xfnm[ R.]}.
\newblock \bibinfo{title}{Brain computer interface for controlling rc-car using
  emotiv epoc+}.
\newblock \bibinfo{journal}{Journal of Telecommunication}
  \bibinfo{year}{2018};\bibinfo{volume}{10}.
%Type = Article
\bibitem[{Kwak et~al.(2017)Kwak, Müller and Lee}]{MultiChannelCNN}
\bibinfo{author}{Kwak\xfnm[ N.S.]}, \bibinfo{author}{Müller\xfnm[ K.R.]},
  \bibinfo{author}{Lee\xfnm[ S.W.]}.
\newblock \bibinfo{title}{A convolutional neural network for steady state
  visual evoked potential classification under ambulatory environment}.
\newblock \bibinfo{journal}{PLOS ONE}
  \bibinfo{year}{2017};\bibinfo{volume}{12}(\bibinfo{number}{2}):\bibinfo{pages}{1--20}.
\newblock \URLprefix \url{https://doi.org/10.1371/journal.pone.0172578}.
  \DOIprefix\doi{10.1371/journal.pone.0172578}.
%Type = Article
\bibitem[{{Nguyen} and {Chung}(2019)}]{SingleChannelCNN}
\bibinfo{author}{{Nguyen}\xfnm[ T.]}, \bibinfo{author}{{Chung}\xfnm[ W.]}.
\newblock \bibinfo{title}{A single-channel ssvep-based bci speller using deep
  learning}.
\newblock \bibinfo{journal}{IEEE Access}
  \bibinfo{year}{2019};\bibinfo{volume}{7}:\bibinfo{pages}{1752--1763}.
\newblock \DOIprefix\doi{10.1109/ACCESS.2018.2886759}.
%Type = Misc
\bibitem[{Lee and Lee(2020)}]{TimeAndFreqCNN}
\bibinfo{author}{Lee\xfnm[ Y.E.]}, \bibinfo{author}{Lee\xfnm[ M.]}.
\newblock \bibinfo{title}{Decoding visual responses based on deep neural
  networks with ear-eeg signals}.
\newblock \bibinfo{year}{2020}.
\newblock \href{http://arxiv.org/abs/2002.01085}{\tt arXiv:2002.01085}.
%Type = Article
\bibitem[{Beverina et~al.(2003)Beverina, Palmas, Silvoni, Piccione, Giove
  et~al.}]{beverina}
\bibinfo{author}{Beverina\xfnm[ F.]}, \bibinfo{author}{Palmas\xfnm[ G.]},
  \bibinfo{author}{Silvoni\xfnm[ S.]}, \bibinfo{author}{Piccione\xfnm[ F.]},
  \bibinfo{author}{Giove\xfnm[ S.]}, et~al.
\newblock \bibinfo{title}{User adaptive bcis: Ssvep and p300 based interfaces.}
\newblock \bibinfo{journal}{PsychNology Journal}
  \bibinfo{year}{2003};\bibinfo{volume}{1}(\bibinfo{number}{4}):\bibinfo{pages}{331--354}.
%Type = Article
\bibitem[{Galloway(1990)}]{galloway1990}
\bibinfo{author}{Galloway\xfnm[ N.]}.
\newblock \bibinfo{title}{Human brain electrophysiology: Evoked potentials and
  evoked magnetic fields in science and medicine}.
\newblock \bibinfo{journal}{The British journal of ophthalmology}
  \bibinfo{year}{1990};\bibinfo{volume}{74}(\bibinfo{number}{4}):\bibinfo{pages}{255}.
%Type = Article
\bibitem[{Bassi et~al.(2019)Bassi, Rampazzo and
  Attux}]{bassi_rampazzo_attux_2019}
\bibinfo{author}{Bassi\xfnm[ P.R.A.S.]}, \bibinfo{author}{Rampazzo\xfnm[ W.]},
  \bibinfo{author}{Attux\xfnm[ R.]}.
\newblock \bibinfo{title}{Redes neurais profundas triplet aplicadas à
  classificação de sinais em interfaces cérebro-computador}.
\newblock \bibinfo{journal}{Anais do XXXVII Simpósio Brasileiro de
  Telecomunicações e Processamento de Sinais} \bibinfo{year}{2019};.
%Type = Article
\bibitem[{{Trunk}(1979)}]{Curse}
\bibinfo{author}{{Trunk}\xfnm[ G.V.]}.
\newblock \bibinfo{title}{A problem of dimensionality: A simple example}.
\newblock \bibinfo{journal}{IEEE Transactions on Pattern Analysis and Machine
  Intelligence}
  \bibinfo{year}{1979};\bibinfo{volume}{PAMI-1}(\bibinfo{number}{3}):\bibinfo{pages}{306--307}.
%Type = Article
\bibitem[{Pan and Yang(2009)}]{pan2009}
\bibinfo{author}{Pan\xfnm[ S.J.]}, \bibinfo{author}{Yang\xfnm[ Q.]}.
\newblock \bibinfo{title}{A survey on transfer learning}.
\newblock \bibinfo{journal}{Transactions on Knowledge and Data Engineering}
  \bibinfo{year}{2009};\bibinfo{volume}{22}(\bibinfo{number}{10}):\bibinfo{pages}{1345--1359}.
%Type = Article
\bibitem[{Bengio et~al.(2013)Bengio, Courville and Vincent}]{bengio2013}
\bibinfo{author}{Bengio\xfnm[ Y.]}, \bibinfo{author}{Courville\xfnm[ A.]},
  \bibinfo{author}{Vincent\xfnm[ P.]}.
\newblock \bibinfo{title}{Representation learning: A review and new
  perspectives}.
\newblock \bibinfo{journal}{Transactions on Pattern Analysis and Machine
  Intelligence}
  \bibinfo{year}{2013};\bibinfo{volume}{35}(\bibinfo{number}{8}):\bibinfo{pages}{1798--1828}.
%Type = Inproceedings
\bibitem[{Yosinski et~al.(2014)Yosinski, Clune, Bengio and
  Lipson}]{yosinski2014}
\bibinfo{author}{Yosinski\xfnm[ J.]}, \bibinfo{author}{Clune\xfnm[ J.]},
  \bibinfo{author}{Bengio\xfnm[ Y.]}, \bibinfo{author}{Lipson\xfnm[ H.]}.
\newblock \bibinfo{title}{How transferable are features in deep neural
  networks?}
\newblock In: \bibinfo{booktitle}{Advances in Neural Information Processing
  Systems}. \bibinfo{year}{2014}, p. \bibinfo{pages}{3320--3328}.
%Type = Misc
\bibitem[{Alu(2019)}]{FBCCADrone}
\bibinfo{author}{Alu\xfnm[ E.]}.
\newblock \bibinfo{title}{Cecnl\_realtimebci}.
\newblock \bibinfo{year}{2019}.
\newblock \URLprefix
  \url{https://github.com/eugeneALU/CECNL_RealTimeBCI/blob/master/fbcca.py}.

\end{thebibliography}

\end{document}